\documentclass[aps,prb,showpacs, showkeys]{revtex4}
\usepackage[english]{babel}
\usepackage[dvips]{epsfig}

\begin{document}
\title{Shape-induced phenomena in the finite-size antiferromagnets}
\author{ Helen V. Gomonay and Vadim M. Loktev}
\affiliation{Bogolyubov Institute for Theoretical Physics NAS of
Ukraine,\\ Metrologichna str. 14-b, 03680, Kyiv, Ukraine}
\begin{abstract}
It is of common knowledge that the direction of easy axis in the
finite-size ferromagnetic sample is controlled by its shape. In
the present paper we show that a similar phenomenon should be
observed in the compensated antiferromagnets with strong
magnetoelastic coupling. Destressing energy which originates from
the long-range magnetoelastic forces is analogous to
demagnetization energy in ferromagnetic materials and is
responsible for the formation of equilibrium domain structure and
 anisotropy of macroscopic magnetic properties. In particular,
crystal shape may be a source of additional uniaxial magnetic
anisotropy which removes degeneracy of antiferromagnetic vector or
artificial 4th order anisotropy in the case of a square
cross-section sample. In a special case of antiferromagnetic
nanopillars shape-induced anisotropy can be substantially enhanced
due to lattice mismatch with the substrate. These effects can be
detected by the magnetic rotational torque and antiferromagnetic
resonance measurements.
\end{abstract}
\maketitle

\section{Introduction}\label{introduction}

The fact that antiferromagnetic crystals  break up into the
regions with different orientation of antiferromagnetic vectors
below the N{\'e}el temperature was predicted theoretically by L.
N{\'e}el \cite{Neel:1953} and then proved experimentally (see,
e.g., \cite{Wilkinson:1959, Tanner:1976, Baruchel:1977,
Janossy:1999, Scholl:2000, Hillebrecht:2001} and many others).

Domain structures observed in different antiferromagnets have some
common features that we summarize below.
\begin{enumerate}\renewcommand{\theenumi}{\roman{enumi}}\renewcommand{\labelenumi}{\theenumi)}
  \item\label{i} Magnetic domains with different
orientations of antiferromagnetic vector are characterized by
different tensors of spontaneous strain and so can be treated as
deformation twins.
  \item\label{ii} The morphology of antiferromagnetic domains is
similar to the morphology of deformation twins in martensites.  In
contrast to ferromagnets, domain structure in antiferromagnets is
regular, periodic and consists of alternating stripes with
different deformation.
  \item\label{iii} Unlike ferromagnets, domain walls separating domains with nonparallel antiferromagnetic vectors  are
plane-like and are parallel to low-index atomic planes.
\item\label{iv}
Deformation does not map the orientation of antiferromagnetic
vector locally (e.g., inside the domain wall orientation of
antiferromagnetic vector is determined by competition between
exchange interaction and deformation-induced anisotropy).
\item \label{v}Antiferromagnetic domains spontaneously appear below the
N{\'e}el temperature. The domain patterns observed during
heating-cooling cycles through the N{\'e}el point may be either
identical or similar to each other.
\item\label{vi} Domain structure may be reversibly
changed by external magnetic field or stress.
\end{enumerate}

The properties (\ref{i})-(\ref{iv}) show that magnetoelastic
coupling plays the leading role in formation of the domain
structure in antiferromagnets. It follows from (\ref{v}),
(\ref{vi}) that the domain structure may be considered as
  thermodynamically equilibrium notwithstanding the fact that formation of the domain walls
is associated with positive contribution into free energy of the
whole sample.
  Regularity of the domain structure (properties (\ref{ii}),
  (\ref{iii})) excludes an entropy of domain disorder as
  a factor which leads to a decrease of free energy of a
  sample and favors formation of inhomogeneous state \cite{Li:1956}.
  Properties (\ref{iv})-(\ref{vi})
  may be explained by presence of the elastic defects (dislocations, disclinations,
  etc.) \cite{Kalita:2004} that produce inhomogeneous stress field in a sample.
This ``frozen-in'' \emph{extraneous} (with respect to an ideal
crystal) field stabilizes inhomogeneous distribution of
antiferromagnetic vector via magnetoelastic interactions and
  ensures reconstruction of the domain structure during heating/cooling
  cycles.

  Another model \cite{gomo:2005(1)}
consistent with all the above mentioned properties is based on the
assumption that antiferromagnetic ordering is accompanied by
appearance of so called quasiplastic stresses \cite{Kleman:1972}
coupled with the orientation of antiferromagnetic vector. We
assume that these \emph{intrinsic} stresses are caused by virtual
forces that represent the change of free energy of the system with
displacement of an atom bearing magnetic moment. Self-consistent
distribution of the internal stress field depends upon the shape
of the sample and is generally inhomogeneous. Equilibrium
distribution of antiferromagnetic vector maps the stress field and
thus is also inhomogeneous and sensitive to application of
external field and temperature variation.

Both the defect-based and defectless models  exploiting
magnetoelastic mechanism predict similar dependence of macroscopic
characteristics of a sample \emph{vs} external magnetic and stress
field, but lead to different results when applied to a set of
different samples. Namely, in the framework of the
\emph{defect-based} model, the domain distribution, domain size,
and some other quantitative characteristics may vary depending on
technological conditions and prehistory of a sample. On the
contrary, the \emph{defectless} model predicts variation of
macroscopic properties of antiferromagnetic crystals with
variation of their shape.

Below we predict shape-related phenomena in antiferromagnets that
can be experimentally tested. In the framework of the
\emph{defectless} model we calculate the effective shape-induced
anisotropy which can be determined by torque measurements, and the
frequency of the lowest spin-wave branch detectable by
antiferromagnetic resonance technique. We consider the case of an
``easy-plane'' antiferromagnet typical example of which is given
by NiO, CoCl$_2$, or KCoF$_3$.

\section{Destressing energy}\label{Destressing}
According to our main assumption, an antiferromagnetic vector
$\mathbf{L}(\mathbf{r})$ may be treated as a quasidefect that
produces intrinsic stress field
$\hat{\sigma}^{(\rm{in})}(\mathbf{r})$. Thus, the thermodynamic
potential of the finite-size antiferromagnet can be represented as
(see, e.g., Ref. \onlinecite{gomo:2005(1)})
\begin{equation}\label{free-energy}
  \Phi[\mathbf{L}(\mathbf{r})]=\int_V\phantom{!}\left\{f^{\rm
{mag}}[\mathbf{L}(\mathbf{r})]-\frac{1}{2}\hat
\sigma^{(\rm{in})}[\mathbf{L}(\mathbf{r})]:\hat c^{-1}:\hat
\sigma^{(\rm{in})}[\mathbf{L}(\mathbf{r})]\right\}\,d\mathbf{r}+
\Phi^{\rm{dest}}\,.
\end{equation}
Here $f^{\rm {mag}}$ is a ``bare'' magnetic energy, the second
term is a self-energy of quasidefect, $\hat{c}^{-1}$ is 4-th rank
tensor of elastic stiffness, $\Phi^{\rm{dest}}$ is a
 destressing energy\cite{gomo:2005(1)} which describes interaction
between the quasidefects localized at different points. Sign ``:''
is used to denote an inner product between the 2nd rank tensors.

Explicit expression for destressing energy is obtained from the
requirement for mechanical equilibrium with due account of
boundary conditions at the sample surface. The main nonnegative
contribution arises from averaged (over the sample volume $V$)
internal stress $\langle\hat\sigma^{\rm
  in}\rangle$ and can be represented as
\begin{equation}\label{Fourier_2}
  \Phi^{\rm dest}=\frac{V}{2}\langle\sigma^{\rm in}_{jl}\rangle\aleph_{jklm}\langle\sigma^{\rm
  in}_{km}\rangle,\quad\aleph_{jklm}\equiv\frac{\partial^2}{\partial r_k\partial
r_m}\int_V G_{jl}(\mathbf{r}-\mathbf{r}')\,d\mathbf{r}',
\end{equation}
where $G_{km}(\mathbf{r}-\mathbf{r}')$ is a 3D Green's function of
elasticity (with zero nonsingular part) and 4-th rank symmetrical
destressing tensor $\hat{\hat{\aleph}}$ depends upon the sample
shape.

 Functional dependence between intrinsic stress tensor and antiferromagnetic
vector is given by a constitutive relation which should satisfy
the principles of locality, material objectivity and material
symmetry. The simplest form of such a relation which assumes
isotropy of magnetoelastic properties of the media is
\begin{equation}\label{intrinsic_stress}
  \sigma^{\rm in}_{jk}=\frac{\lambda_v}{3}\mathbf{L}^2\delta_{jk}+\lambda^\prime
  \left(L_jL_k-\frac{\mathbf{L}^2}{3}\delta_{jk}\right),
\end{equation}
where coefficients $\lambda_v$ and $\lambda^\prime$ define the
principal stresses of magnetoelastic nature.

Substituting (\ref{Fourier_2}) and (\ref{intrinsic_stress}) into
(\ref{free-energy}) one comes to a closed expression for
thermodynamic potential, minimization of which gives equilibrium
distribution of $\mathbf{L}$ throughout the sample.

Two terms of magnetoelastic origin in (\ref{free-energy}) have one
principal distinction. The structure of the local energy
contribution (second term) is defined by crystal symmetry, while
the structure of $\Phi^{\rm dest}$ depends upon the sample shape.
In the framework of phenomenological approach local energy
contributes to effective anisotropy constant only, while
destressing energy is responsible for the domain structure
formation and may be a source of artificial 4th order anisotropy
as will be shown below.
\section{Application to an ``easy-plane'' antiferromagnet}\label{Application}
To understand the role of destressing energy in the shape-induced
phenomena we consider the simplest case of an ``easy-plane''
antiferromagnet (point symmetry group of the crystal includes
3-rd, 4-th or 6-th order rotations around $Z$-axis) cut in a form
of an elliptic cylinder with  $a$  and $b$ semiaxes (parallel to
$X$ and $Y$ axes, respectively) and generatrix parallel to $Z$.
 The elastic properties of the media are supposed
to be isotropic ($c_{11}-c_{12}=2c_{44}$). In this case nontrivial
contribution to destressing energy takes a form
\begin{eqnarray}\label{destress_cylinder_2}
  \Phi^{\rm dest}&=&\frac{V}{2}\left\{K^{\rm elas}_{2} (L_Y^2-L_X^2)
+K^{\rm elas}_{\rm is}[\langle
  L_X^2-L_Y^2\rangle^2\right.\nonumber\\
  &+&\left.4\langle
L_XL_Y\rangle^2]-K^{\rm elas}_{\rm 4an}[\langle
  L_X^2-L_Y^2\rangle^2-4\langle
L_XL_Y\rangle^2]\right\},
\end{eqnarray}
where effective shape-induced anisotropy constants are
\begin{eqnarray}\label{destress_constants}
K^{\rm
elas}_{2}&=&\frac{a-b}{a+b}\frac{(\lambda^\prime)^2(2-3\nu)+\lambda_v\lambda^\prime}{4c_{44}(1-\nu)}\,,\nonumber\\
K^{\rm elas}_{\rm
is}&=&\frac{(\lambda^\prime)^2(3-4\nu)}{8c_{44}(1-\nu)}\,, \quad
K^{\rm elas}_{\rm
4an}=\left(\frac{a-b}{a+b}\right)^2\frac{(\lambda^\prime)^2}{6c_{44}(1-\nu)}\,,
\end{eqnarray}
and $\nu=c_{12}/(c_{11}+c_{12})$ is the Poisson ratio.

Magnetic  energy density of such an antiferromagnet in the
external magnetic field $\mathbf{H}$ (low compared with spin-flip
value) may be written as:
\begin{equation}\label{magnetic_energy}
 f^{\rm
{mag}}=\frac{1}{2}K^{\rm mag}_{2}L_Z^2+f^{\rm mag}_{\rm
in-plane}-\frac{1}{2}\chi[\mathbf{H}\times\mathbf{L}]^2\,,
\end{equation}
where $\chi$ is magnetic susceptibility, out-of-plane anisotropy
constant $K^{\rm mag}_{2}\gg K^{\rm elas}_{2}$ is large enough to
keep antiferromagnetic vector in $XY$ plane, and explicit form of
the in-plane magnetic anisotropy $f^{\rm mag}_{\rm in-plane}$ is
specified by a crystal symmetry.

For a typical ``easy-plane'' antiferromagnet $f^{\rm mag}_{\rm
in-plane}$ is much less than the effective constants
(\ref{destress_constants}) of magnetoelastic nature (see
table~\ref{t.1}). So, in such a crystal  destressing effects may
stimulate formation of the domain structure and change equilibrium
orientation of antiferromagnetic vector.

\begin{table}
\caption{Shape-induced $K^{\rm elas}_{\rm is}$ and magnetic
$f^{\rm mag}_{\rm in-plane}$ anisotropy (in erg/cm$^3$), and
critical aspect ratio for typical ``easy-plane'' antiferromagnets
(details of calculation see in \cite{gomo:2002}). } \label{t.1}
\begin{center}
\begin{tabular}{cccc}
Crystal &$K^{\rm elas}_{\rm is}$ &$f^{\rm mag}_{\rm
in-plane}$&$(a/b)_{\rm cr}$ \\ NiO&0.8$\cdot 10^{4}$&288&1.1\\
CoCl$_2$&5.6$\cdot 10^5$& $<3\cdot 10^4$&3.4\\ KCoF$_3$&3$\cdot
10^6$&5$\cdot 10^5$&1.5
\end{tabular}
\end{center}
\end{table}

It should also be stressed that three different shape-induced
anisotropy constants (\ref{destress_constants}) depend on the
aspect ratio $a/b$ in a different way. This opens a possibility to
control macroscopic properties of the sample varying its shape.

\section{Formation of the equilibrium domain structure}
 If  in-plane magnetic anisotropy is small but not
vanishing, then, the destressing energy favors formation of the
domain structure. For example, in the case of an isotropic sample
($a=b$)
 the only nontrivial term with $K^{\rm elas}_{\rm is}$ in eq.~(\ref{destress_cylinder_2}) is
nonnegative. In the absence of external field it can only be
diminished by zeroing average values of $\langle
  L_x^2-L_y^2\rangle$, $\langle
  L_xL_y\rangle$, i.e., by appearance of equiprobable
  distribution of domains with different orientation of antiferromagnetic vector.

The external magnetic field causes rotation of  antiferromagnetic
vector  and removes degeneracy of various domains. Due to a
long-range character of elastic forces, the field-induced
ponderomotive force that acts on the domain wall is compensated by
the destressing, restoring force. Competition of these two factors
determines equilibrium proportion of different domains. If, for
example, external field is applied  in-parallel to an ``easy''
direction in $XY$ plane (say, $X$ axis), then, the volume fraction
$\xi$ of energetically preferable $Y$-type domain (in which
$\mathbf{L}\perp\mathbf{H}$), increases quadratically with field
$\xi=0.5[1+(H/H_{\rm{MD}})^2]$ up to monodomainization field
$H_{\rm{MD}} = \sqrt {K_{\rm{is}}^{\rm{elas}} / \chi } \propto
\lambda ' / \sqrt {\chi c_{44} }$.


The domain structure may be also observed in samples with small
but nonzero eccentricity ($a \approx b$) providing that in-plane
magnetic anisotropy is large enough to keep two different
equilibrium (stable and metastable) $\mathbf{L}$ orientations:
 $K_2^{\rm{elas}} \le
f_{\rm{in - plane}}^{\rm{mag}} $. In this case it is the shape
factor that removes degeneracy  and hence equiprobability of the
domains. In the above example the fraction of $X$-type domain
($\mathbf{L}\|X$) depends on the aspect ratio as follows
\begin{equation}
\label{fraction} \xi - \frac{1}{2} = \frac{K_2^{\rm{elas}}
}{K_{4\rm{an}}^{\rm{elas}} } \propto \frac{b - a}{b + a}
\end{equation}
and  the domain structure reproduces the orthorhombic symmetry of
the sample.

Critical values of aspect ratio $(a/b)_{\rm cr}$ (obtained from
the condition  $K_{2}^{\rm{elas}} = f_{\rm{in -
plane}}^{\rm{mag}}$) at which equilibrium domain structure is
still thermodynamically favorable are given in the last column of
table~\ref{t.1}.
\section{Torque effect}
If the aspect ratio $a/b$ of the sample noticeably differs from 1,
then, all the effective anisotropy constants in
(\ref{destress_cylinder_2}) have the same order of value and are
much greater than in-plane anisotropy, $K_{\rm
{is}}^{\rm{elas}}\gg f_{\rm{in-plane}}^{\rm{mag}}$ (see
table~\ref{t.1}). So, the sample has shape-induced uniaxial
anisotropy regardless of its crystallographic symmetry.

\begin{figure}[htbp]
\includegraphics[scale=0.5]{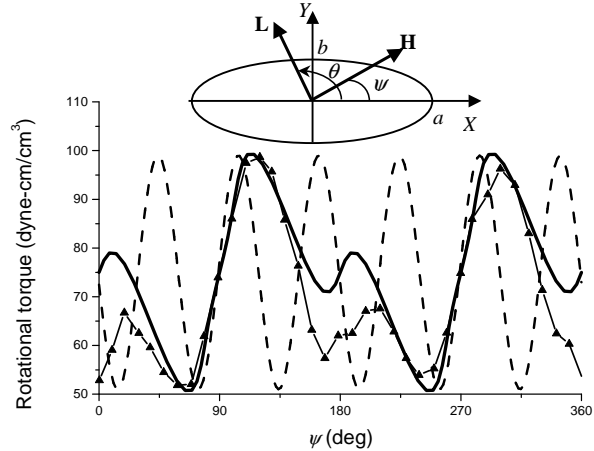} \caption{Rotational
torque of untwinned NiO crystal at RT in H=4.8 kOe for [111]
rotational axis. Triangles - experimental data
\cite{Roth:1960(2)}, solid line -- theoretical approximation
according to (\ref{rotational_torque}), (\ref{minimum_angle}),
dashed line -- approximation of infinite crystal (no shape
effect).} \label{fig_2}
\end{figure}

An appropriate tool for measuring anisotropy constants is a
rotational torque of untwinned crystal in the magnetic field. If
the rotational axis is perpendicular to ``easy-plane'' $XY$ and
magnetic field makes a $\psi$ angle with $X$ axis (see inset in
fig.~\ref{fig_2}), then the rotational torque can be calculated as
$-\partial \Phi/\partial \psi$ where free energy potential $\Phi$
is given by eq.~(\ref{free-energy}). With account of the
eqs.~(\ref{destress_cylinder_2})-(\ref{magnetic_energy}) the
rotational torque per unit volume is represented as
\begin{equation}\label{rotational_torque}
T(\psi)=K^{\rm elas}_{\rm 2}\sin 2\theta(\psi)+2K^{\rm elas}_{\rm
4an}\sin
  4\theta(\psi),
\end{equation}
where a $\theta$ angle between antiferromagnetic vector and $X$
axis
 unambiguously determines equilibrium orientation of $\mathbf{L}$ and
is calculated from the condition for minimum of the potential
(\ref{free-energy}):
\begin{equation}\label{minimum_angle}
  K^{\rm elas}_{\rm 2}\sin 2\theta+2K^{\rm elas}_{\rm 4an}\sin
  4\theta-\frac{1}{2}\chi H^2\sin 2(\theta-\psi)=0\,.
\end{equation}
Analysis of eqs.~(\ref{rotational_torque})-(\ref{minimum_angle})
shows that i) effective anisotropy is determined by
shape-dependent (via $a/b$ ratio) constants $K^{\rm elas}_{\rm
2}$, $K^{\rm elas}_{\rm 4an}$; ii) shape-induced anisotropy
removes multiaxial degeneracy of equilibrium orientation of
antiferromagnetic vector and thus excludes formation of domain
structure; iii) in the field absence the preferred orientation of
antiferromagnetic vector coincides with the longer ellipse axis
($K^{\rm elas}_{\rm 2}>0$, $a>b$); iv) external magnetic field
applied along the longer ellipse axis may induce spin-flop
transition at $H=H_{\rm{SF}}$, where spin-flop field
\begin{equation}\label{spin-flop}
H_{\rm{SF}} = \sqrt {\frac{1}{ \chi }\left| {K_2^{\rm{elas}} -
4K_{4\rm{an}}^{\rm{elas}} } \right| } \propto
{\frac{\lambda^\prime}{ \sqrt {\chi c_{44} } }} \sqrt {\frac{a -
b} { a + b}}
\end{equation}
is governed by the sample shape and is independent of crystalline
anisotropy.

An interesting result is obtained for the sample having a square
cross-section. In this case equilibrium distribution of
antiferromagnetic vector is in principle inhomogeneous but in
average can still be described by eq.~(\ref{minimum_angle}) with
\begin{equation}\label{destress_constants_square}
K^{\rm elas}_{2}=0, \qquad K^{\rm elas}_{\rm
4an}=\frac{\ln2(\lambda^\prime)^2}{\pi c_{44}(1-\nu)}>0\,.
\end{equation}
It is obvious that such a sample shows 4-th order effective
anisotropy with ``easy axes'' directed along square diagonal
(45$^\circ$ angle with respect to $X$ axis, as results from
condition $K^{\rm elas}_{\rm 4an}>0$).

\begin{figure}\begin{minipage}[b]{0.45\linewidth}
{\centerline{\includegraphics[width=1\textwidth]{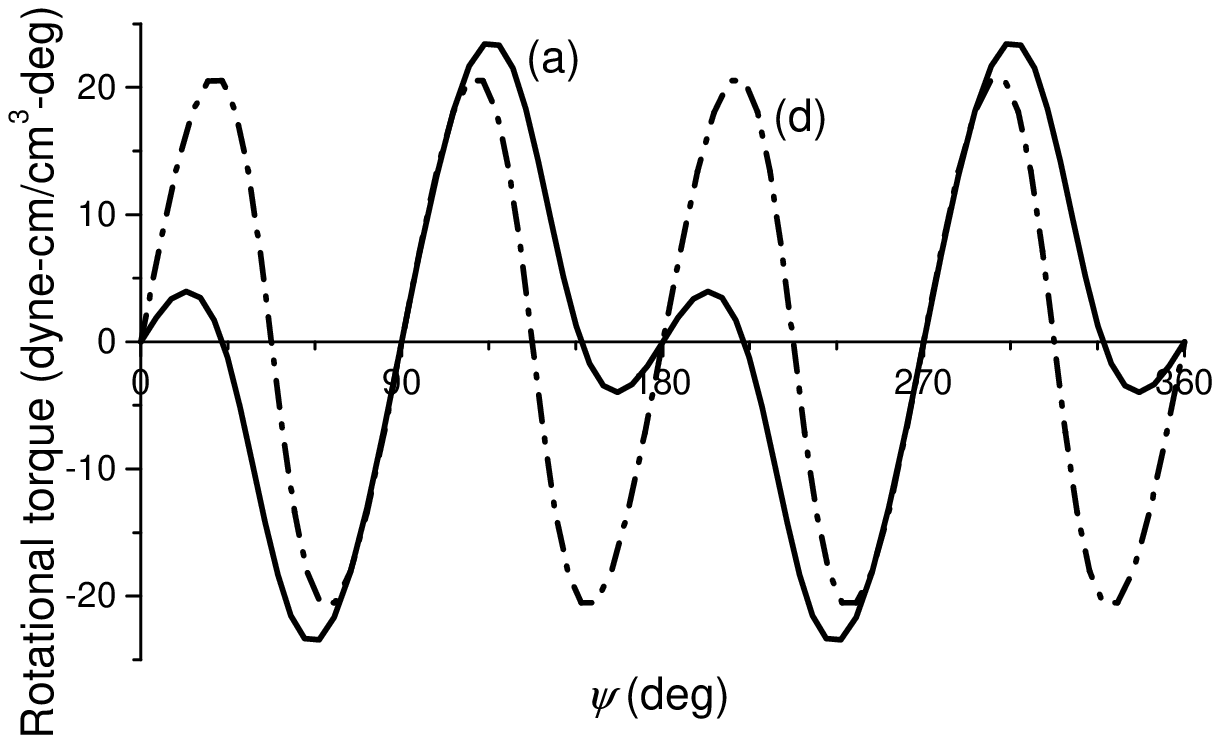}}
}\end{minipage}\hfill
\begin{minipage}[b]{0.45\linewidth}
{\centerline{\includegraphics[width=1\textwidth]{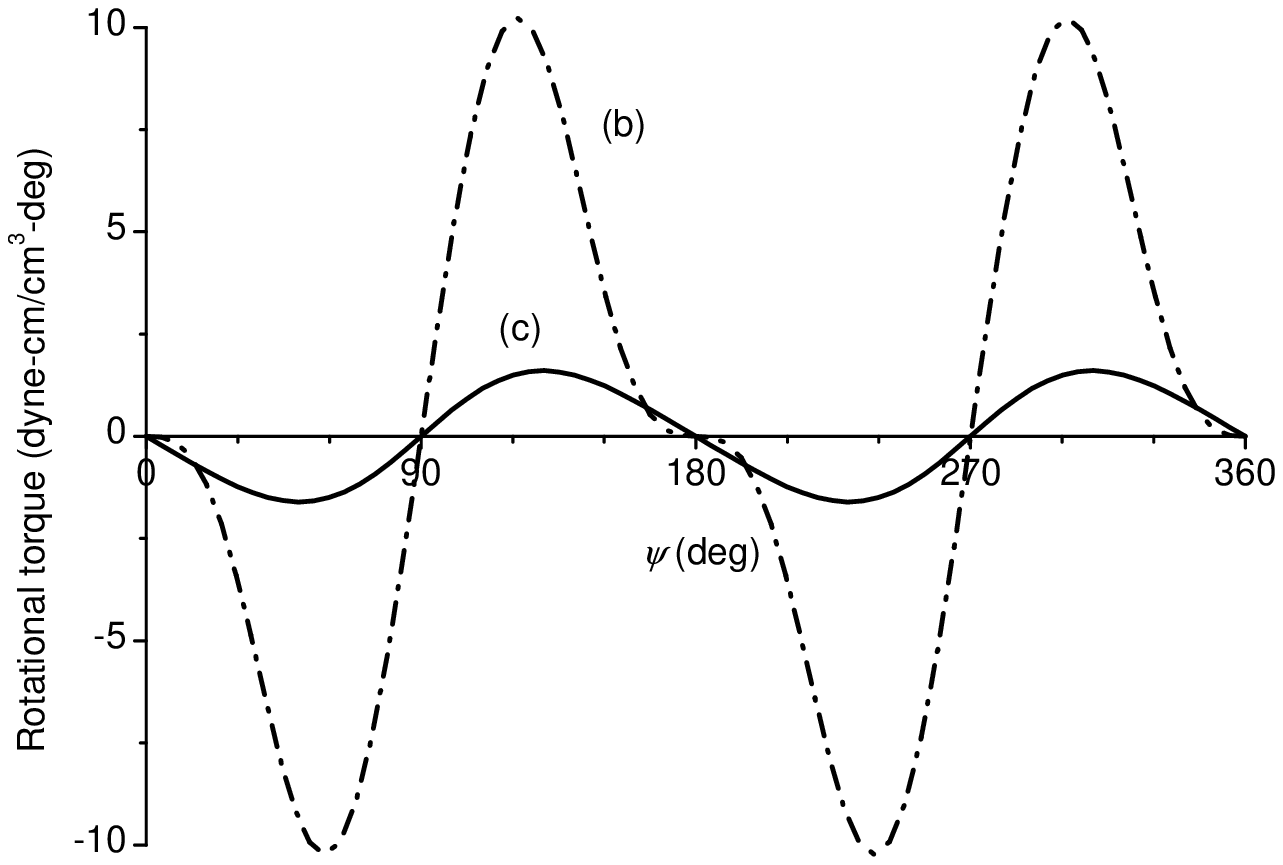}}
}\end{minipage} \caption{Rotational torque (calculated) of
untwinned NiO crystal at RT in H=5 kOe for [111] rotational axis
for the samples with different shape. For ellipse-section samples
an aspect ratio is $a/b=20$ (a); 3 (b); 1.2 (c); (d) --
square-section sample.} \label{fig_3}
\end{figure}

The rotational torque calculated from
eqs.~(\ref{rotational_torque}), (\ref{minimum_angle})
 for a typical antiferromagnet NiO is shown in
figs.~\ref{fig_2}, \ref{fig_3}. In calculations we use
experimental results \cite{Roth:1960(2)} for a rotational torque
around [111] axis taken at room temperature for $H$=4.8 kOe
(triangles in fig.~\ref{fig_2}). Since the experiment shows strong
hysteresis in clock-wise and counter-clock wise rotations, the
data in fig.~\ref{fig_2} are preliminarily averaged over cc-ccw
cycle. Theoretical curve (solid line) includes some nonzero
average torque that may result from inhomogeneity of the sample,
and adjusting parameters are $\chi =4.35 \cdot 10^{ -
4}$~emu/cm$^{3}$, $K_2^{\rm{elas}}=900$~erg/cm$^{3}$,
$K_{4\rm{an}}^{\rm{elas}}=450$~erg/cm$^3$, $a/b=20$.

Four curves in fig.~\ref{fig_3} demonstrate possible variation of
the rotational torque with the crystal shape. For a large aspect
ratio (curve (a)) contributions from both 2-nd and 4-th order
anisotropy terms are equally important ($K_2^{\rm{elas}}\propto
K_{4\rm{an}}^{\rm{elas}}$), the torque curve is composed of
$\sin2\theta$ and $\sin 4\theta$ components. At lower aspect ratio
(curves (b), (c)) the 4-th order component becomes less pronounced
and the amplitude of torque also diminishes.  A circular cylinder
($a=b$) will show no shape effect in a single domain state, but a
sample with a square cross-section should posses 4-th order
anisotropy (as seen from curve (d)) regardless of crystalline
symmetry. 

Evidently, shape-dependence of the magnetic rotational torque for
NiO crystal was observed in Ref.~\onlinecite{Roth:1960(2)}. The
authors notice that ``a nearly pure $\sin 4\theta $ curve is
obtained when the (111) cross section is square''. For an
arbitrary shaped section the experimental curve (fig.~\ref{fig_2},
triangles) is satisfactorily fitted with combination of
$\sin2\theta$ and $\sin 4\theta$ components (solid line). Dashed
line in fig.~\ref{fig_2} shows theoretical curve with $f_{\rm{in -
plane}}^{\rm{mag}}=220\cos 6\theta$~erg/cm$^3$ that could be
expected in neglecting of shape-induced effects. The presence of
$\sin 4\theta$ component in the rotational torque makes it
possible to exclude the effect of $\mathbf{L}$ ``freezing'' by
magnetoelastic strain which may be expected in relatively small
magnetic field. Additional anisotropy induced by frozen lattice is
uniaxial and should be insensible to the variation of a crystal
shape.

\section{Antiferromagnetic resonance} 
\begin{figure}[htbp]
\includegraphics[scale=0.8]{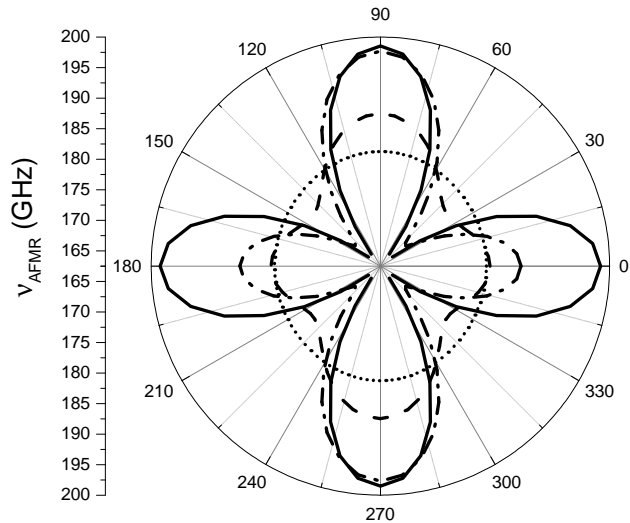} \caption{Angular
dependence of AFMR frequency vs magnetic field orientation for
samples with different aspect ratio: $a/b=20$ (dash-dotted line);
3 (dashed line); 1.2 (dotted line). Solid line corresponds to a
square-section sample.} \label{fig_4}
\end{figure}

The effect of shape-induced anisotropy may also be detected
by measuring frequency of the lowest branch of a spin-wave
spectrum.  Antiferromagnetic resonance (AFMR) frequency calculated
within a standard Lagrangian technique in long-wave approximation
is given by the expression
\begin{equation}
\label{frequency} \nu_{\rm AFMR} = g\sqrt{
\frac{2}{\chi}\left(K_{\rm{is}}^{\rm{elas}} + {K_2^{\rm{elas}}
\cos 2\theta + 4K_{4\rm{an}}^{\rm{elas}} \cos 4\theta - \chi
H^2\cos 2(\theta - \psi )} \right)},
\end{equation}
where $g$ is gyromagnetic ratio and equilibrium value of $\theta$
may be calculated from (\ref{minimum_angle}). Polar diagram of
$\nu_{\rm AFMR}(\psi)$ calculated from (\ref{frequency})  for NiO
($g=2,5$) for a different crystal shape is shown in
fig.~\ref{fig_4}. For the aspect ratio close to 1 (dotted line)
the shape effect is negligible and magnetoelastic gap in AFMR
spectrum is almost isotropic. For elongated (dash-dotted and
dashed lines) or square-shaped (solid line) samples the AFMR gap
should show strong 2-fold or 4-fold anisotropy.

\section{Shape effect in antiferromagnetic nanopillars}\label{sec:nanopillar}
Taking into account recent interest to multilayered structures
based on antiferromagnetic materials we consider possible shape
effects in antiferromagnetic nanopillar that can be a constitutive
part of spin-valve structure (see, e.g. \cite{Urazhdin:2007}).
Typical nanopillar has a form of a very thin (thickness $c\approx
3\div 10$~nm) elliptic cylinder with the pronounced in-plane
aspect ratio (with $a\propto 120$~nm and $b\propto 50$~nm).

In this case ($c\ll a,b$) the constants of shape-induced
anisotropy may be expressed through the parameter $k^2=1-b^2/a^2$
that depends on the aspect ratio $b/a\le 1$, namely,
\begin{equation}\label{destress_constants_thin_film}
K^{\rm
elas}_{2}=\frac{c}{b}\cdot\frac{[(\lambda^\prime)^2(2-3\nu)+\lambda_v\lambda^\prime]J_2(k)}{4c_{44}(1-\nu)},\,
K^{\rm elas}_{\rm
4an}=\frac{c}{b}\cdot\frac{2(\lambda^\prime)^2J_4(k)}{3c_{44}(1-\nu)}\,,
\end{equation}
where we have introduced the dimensionless shape-factors $J_{2,4}$
as follows
\begin{eqnarray}\label{constants}
  J_2(k)&=&\int_0^{\pi/2}\frac{(\sin^2\phi+\cos2\phi/k^2)d\phi}{\sqrt{1-k^2\sin^2\phi}},\nonumber\\
J_4(k)&=&
\int_0^{\pi/2}\frac{(1-8\cos2\phi-k^2\sin^2\phi+8\cos2\phi/k^2)d\phi}{\sqrt{1-k^2\sin^2\phi}}.
\end{eqnarray}
\begin{figure}[htbp]
\includegraphics[scale=0.4]{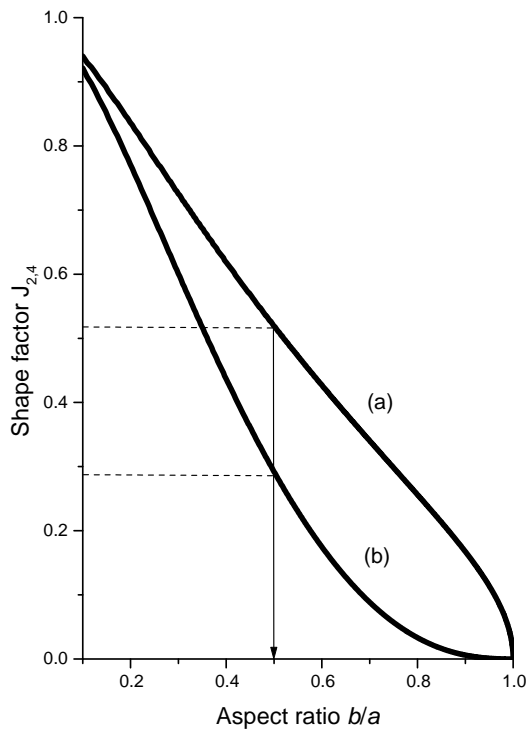}
\caption{The 2-nd (a) and 4-th order (b) shape-induced anisotropy
of thin nanopillar as a function of aspect ratio. Arrow indicates
the typical value of $b/a$ used in the experiments
\cite{Urazhdin:2007}.}\label{fig_5}
\end{figure}
Dependence of the shape-factors $J_{2,4}$ vs aspect ratio
calculated according to eq.~(\ref{constants}) is given in
Fig.\ref{fig_5}. It can be easily understood that both
shape-induced constants (\ref{destress_constants_thin_film})
vanish for isotropic sample ($b=a$). For all the values of aspect
ratio the 2-nd order term is greater than that of the 4-th order,
$J_2\ge J_4$. In the experimentally accessible range of values
$b\propto 0.5 a$ $J_2=0.52$ and $J_4=0.3$. Thus, characteristic
value of the shape-induced anisotropy in the stress-free thin film
can be of the order magnetoelastic energy
$K_{\rm{is}}^{\rm{elas}}\propto(\lambda^\prime)^2/c_{44}$
multiplied by small factor $c/b$ that varies within the range
0.05$\div$0.3 depending on the film thickness.

Substantial enhancement of the 2nd order shape-induced anisotropy
$K^{\rm elas}_{2}$ should be expected in the case of mismatch
between antiferromagnet and substrate lattices. Lattice misfit is
a source of rather strong (usually isotropic) in-plane stresses
$\sigma^{\rm mf}_{xx}=\sigma^{\rm mf}_{yy}=\sigma^{\rm mf}$ that
should be added to the intrinsic stresses
(\ref{intrinsic_stress}). Corresponding (and principal)
contribution into the effective 2-nd order anisotropy constant
takes a form
\begin{equation}\label{stress_2_nd_order}
  K^{\rm
elas}_{2}=\frac{c}{b}\cdot\frac{\sigma^{\rm mf}\lambda^\prime
J_2(k)}{4c_{44}(1-\nu)}.
\end{equation}
If $u^{\rm mf}$ is lattice mismatch and $u^{\rm
spon}\propto\lambda^\prime/c_{44}$ is an observable spontaneous
strain that occurs in the N{\'e}el point, then, we can estimate
$\sigma^{\rm mf}\propto c_{44}u^{\rm mf}$ (in assumption that the
elastic modula of substrate and antiferromagnet are of the same
order of value), $K_{\rm{is}}^{\rm{elas}}\propto
(\lambda^\prime)^2/c_{44}$ and hence,
\[K^{\rm
elas}_{2}\propto \frac{c}{b}\frac{u^{\rm mf}}{u^{\rm
spon}}K_{\rm{is}}^{\rm{elas}}.\] Substituting typical values of
\emph{small} lattice misfit  $u^{\rm mf}=0.005$ and \emph{large}
spontaneous striction $u^{\rm spon}=10^{-4}$ we see that even for
very thin nanopillars with $c/b=0.05$ shape induced anisotropy may
be as large as $K^{\rm elas}_{2}=2.5K_{\rm{is}}^{\rm{elas}}$ and
thus much greater than the ``bare'' in-plane magnetic anisotropy
of antiferromagnet (see table \ref{t.1}).

Nontrivial relation (\ref{stress_2_nd_order}) between the shape of
the sample and external stress produced by the substrate may
reveal itself in switching of the shape-induced direction of easy
axis for different substrates. Really, if $K^{\rm elas}_{2}>0$,
then, equilibrium orientation of $\mathbf{L}$ in monodomain sample
is parallel to the ellipse's long axis $a$ ($X$-direction), as can
be seen from (\ref{destress_cylinder_2}) and inset in
fig.~\ref{fig_2}. According to equation (\ref{stress_2_nd_order}),
the sign of $K^{\rm elas}_{2}$ depends upon the relation between
intrinsic ($\lambda^\prime, u^{\rm spon}$) and extrinsic
($\sigma^{\rm mf}, u^{\rm mf}$) stresses (or strains). In the case
when both substrate and eigen magnetoelastic forces  of
antiferromagnet ``work'' in the same direction, trying to extend
(compress) crystal lattice, product $\sigma^{\rm
mf}\lambda^\prime$ will be positive (in accordance with Le
Chatelier principle) and $K^{\rm elas}_{2}>0$. If we use another
substrate which produces misfit of opposite sign, $K^{\rm
elas}_{2}<0$ and equilibrium orientation of $\mathbf{L}$ will be
parallel to the short axis $b$ ($Y$-direction).

Controlling of spin-orientation by substrate-induced strain was
recently observed\cite{Csisza:2005} in antiferromagnet CoO. In
bulk samples CoO is compressed in $\mathbf{L}$ direction. When
grown on Mg(100) substrate, CoO lattice is expanded in plane and
experiment shows that Co spin go out of plane. And on contrary,
in-plane ordering is observed for Ag(100) substrate, which
produces slight contraction in the film plane. From our point of
view, analogous experiments with nanopillars can be very
instructive in further study of shape-induced effect in
antiferromagnetic crystals.

\section{Conclusions}
\label{sec:mylabel1}

In summary, we propose a model that describes an antiferromagnet
with the pronounced magnetoelastic coupling. The model is based on
the assumption that antiferromagnetic ordering is accompanied by
appearance of elastic dipoles. Due to the long-range nature of
elastic forces, the energy of dipole-dipole interaction
(destressing energy) in the finite-size sample depends on the
crystal shape and is proportional to its volume.

In the ``easy-plane'' antiferromagnets with degenerated
orientation of easy axis the destressing effects may stimulate
formation of the domain structure and re-distribution of the
domains in the presence of external magnetic field.

The model predicts existence of the shape-induced magnetic
anisotropy which corresponds to macroscopic symmetry of the sample
and can be detected by the magnetic rotational torque and AFMR
measurements.

Crystal shape may be a source of additional magnetic uniaxial
anisotropy which produces different effects depending of aspect
ratio. Below some critical value of $a/b$ shape-induced anisotropy
favors formation of the domain structure even in the absence of
any external field. For large aspect ratio (above the critical
value) shape-induced anisotropy removes degeneracy of easy-axis in
a single domain sample. Energy difference between thus induced
easy- and hard-directions depends upon $a/b$. Square cross section
sample should acquire 4th order anisotropy (irrespective to the
crystal symmetry in easy plane). This opens a possibility to
control macroscopic properties of the sample varying its shape.

The shape of antiferromagnetic nanopillars imbedded into
structures with lattice mismatch may be a principal source of
magnetic anisotropy. This fact should be taken into account in
engineering of spin-valve devices.



\end{document}